\documentclass[12pt]{article}
\usepackage{amssymb}
\usepackage{hyperref}
\usepackage{epsfig}
\newtheorem{theorem}{\sc Theorem}[section]
\newtheorem{definition}{\sc Definition}[section]

\newtheorem{Ack}{\it Acknowledgements}[section]
\begin{document}
\baselineskip=24 pt 
\title{\bf A ptototype compartmental model of blood pressure distribution}

\author{M. U. Akhmet$^1$\thanks{Corresponding author. M.U. Akhmet is previously known as M. U. Akhmetov.}, 
 G.A. Bekmukhambetova$^2$}
\date{{\small$^1$ Department of Mathematics, Middle East
Technical University, 06531 Ankara, Turkey, marat@metu.edu.tr \\$^2$ West-Kazakhstan State Medical Academy named after M. Ospanov ,}\\ {\small Mares'eva str.,
68, 463000 Aktobe/Kazakhstan}}
\maketitle 
\vspace{0.5cm}


\begin{abstract}
We consider a system of differential equations the behavior of which solutions possesses several properties characteristic of the blood pressure distribution. The system can be used for a compartmental modeling  \cite{bellman,j} of the cardiovascular system. It admits a unique bounded  solution such that  all coordinates  of the solution are separated from zero by  positive numbers, and  which is  periodic, eventually periodic  or  almost periodic  depending on the moments of heart contraction.   Appropriate numerical simulations are provided.  
\end{abstract}
\maketitle

\section{Introduction and Preliminaries}

A living organism  can be considered as  a complex system of nonlinear oscillating structures of different origins. The system as an integrity  is a "constellation" of all of its oscillations. The
    oscillators  are  connected with each other to  build  oscillating chains.
    The essential task of controlling processes in a living organism is to sustain the oscillating activity of certain significantly unstable elements in such a way, that the amplitudes of their oscillations don't become excessively large or excessively small. It is, therefore,  natural to assume that the data  of oscillations should be  considered as one of the most important sources of information about the condition of the organism.
  
   The movement of any fluid in a certain direction is the result of pressure differences in the starting and ending points of the system. The difference of pressure in heart chambers and in aorta, in aorta and outgoing blood vessels is the cause of blood flow in an organism. 
   Last decades there has been a rise of intensive investigations of blood pressure in its connections with other parameters of the cardiovascular system  \cite{ab,akhmet1},\cite{arnold}-\cite{glass},\cite{hop,lerma,Mc},\cite{nic}-\cite{ped},\cite{ste}.
   At the same time we should recognize that due to the nonlinear properties of blood vessel walls and their variability along the length of a vessel the  relation between actual oscillations of pressure and of blood volume \cite{fung},\cite{Mc},\cite{ped} is  very complex, and consequently,  it is reasonable to consider the distribution of blood  pressure out of it's connection with other parameters, particularly blood volume. 
     There are many different approaches to the problem of modeling cardiovascular dynamics
including, for example, modeling of the coupling mechanisms as a system of differential equations with delay \cite{cav,g} or as a system of ordinary differential equations  \cite{hop,lerma,ste}.
        Compartmental models have been effectively applied to investigate different type processes in
    chemistry \cite{nicolis},  medicine  \cite{And,g,j,ladde}, epidemiology \cite{mur}, ecology \cite{matis}, pharmacokinetics \cite{And,bellman,res,sol}. 
The idea of compartments makes our investigation more abstract. Particularly,  we do not specify the physiological meaning of the   compartments nor do we specify the number of these parts.  Only  compartments which have an essential influence on the aortic  pressure are considered. 

     It is obvious  that all parts of the cardiovascular system  should be involved  if, let say, the blood flow problem is investigated, because the total volume of the blood is constant in the time,  but the pressure  could be discussed locally, as, for example, in \cite{hop,lerma}, where blood pressure for only  compartment is studied.   In our paper we focus on the interactions of the blood pressures of several  parts of the system, but not all of them.  
We suppose that there exists, as a consequence of the blood float, a mutual interaction  of blood pressures in different parts of the system. One of them, the systemic arterial pressure, should be singled out, as it is disturbed impulsively with large simultaneous inflow of blood in the time of heart contraction.
Other  compartments of the system do not directly undergo the impulsive action, so their change is continuous. We want to note that in this paper we consider exclusively the aorta and the parts of the cardiovascular system interacting with the aorta through blood pressure, that is, the compartments belonging to the systemic circulation only are considered. The blood pressure in pulmonary artery also changes impulsively, and the differential equations system we propose can therefore be also used to model the pressure in this artery and its neighboring compartments. We, however, leave it out of our discussion.  
  
 The  goal of our paper is not to  build a final model of cardiovascular system, but rather to introduce a system of differential equations, in which one coordinate is perturbed impulsively and other coordinates exhibit a continuous change,  thus having certain properties characteristic of the actual blood pressure behavior. The obtained  results now need qualitative experimental confirmation.   We hope that our results will serve as a basis for further experimental and clinical studies,  which can lead to better understanding of the cardiovascular system.

Let $\mathbb Z, \mathbb N$ and $\mathbb R$ be the sets of all integers, natural  and real
numbers, respectively, $\mathbb R_+ = [0, \infty),$  and let $||\cdot||$ be the Euclidean norm in
$\mathbb R^n$, $n \in \mathbb N.$
 
  We denote  $C_0$ the aorta and arteries, and   $P_0$  blood pressure in $C_0.$ Let  $C_i,i  = \overline{1,m},$
be other vessels and organs, which adjoin to $C_0,$ and have an essential influence on fluctuations of 
 $P_0,$ and  $ P_i,i  = \overline{1,m},$  are the blood pressure values  in  compartments $C_i,\overline{1,m}.$ 

In our paper we suppose that the pressure variables satisfy the following system of differential equations with  impulses
\begin{eqnarray}
	&&\frac{dP_0(t)}{dt} =  - k_0 P_0  - g_0(P_0-P_1,P_0-P_2,\ldots,P_0-P_m),\nonumber\\
	&&\frac{dP_1(t)}{dt} =  - k_1 P_1 +g_1(P_0-P_1),\nonumber\\	
	&&\frac{dP_2(t)}{dt} =  - k_2 P_2  + g_2(P_0-P_2),\nonumber\\
	&&\ldots \ldots \nonumber\\
	&&\frac{dP_m(t)}{dt} = - k_m P_m  + g_m(P_0-P_m), \nonumber\\
	&& \Delta P_0 |_{t=\theta_i} = I_0  + J_0(P_0),
		\label{1}
\end{eqnarray}	
where $\Delta P_0 |_{t=\theta_i}\equiv P_0(\theta_i+) -P_0(\theta_i), P_0(\theta_i+)= \lim_{t \to \theta_i+} P_0 (t).$

The following assumptions for (\ref{1}) will be needed throughout the paper:
\begin{itemize}
\item[(C1)] real constants  $I_0, k_i, i  = \overline{0,m},$ are positive;
\item[(C2)]  $J_0, g_i$ are real valued functions, $J_0(0) = 0, g_i(0) =0, i= \overline{1,m},J_0(z) > 0, g_i(z)> 0,$ if $z >0, \, g_0(0,0,\ldots,0) = 0,\,g_0(z_1,z_2,\ldots,z_m)> 0,$ if $z_j >0, j = \overline{1,m};$ 
\item[(C3)]  the functions $J_0, g_i, \overline{0,m},$ satisfy the Lipschitz condition 
\begin{eqnarray}
	&& |g_i(z^1) -g_i(z^2)| \le  l_i|z^1 -z^2|, i = \overline{1,m}, \nonumber\\
	&& |g_0(z_1,\ldots,z_m) -g_0(w_1,\ldots,w_m)| \le  l_0 \sqrt{ \sum\limits_{i = \overline{1,m}} |z_i - w_i|^2}, \nonumber\\ 
	&& |J_0(z^1)-J_0(z^1)| \le l_J|z^1 -z^2|;
\label{top-top}  
\end{eqnarray}
\item[(C4)] there exists a number $\omega >0,$ such that $i \omega \le \theta_i < (i+1)\omega, i \in \mathbb Z;$
\item [(C5)]  there exist positive real constants $m_i, i = \overline{0,m}, m_J$ such that
	
\begin{eqnarray}
	&&  \sup\limits_{z \ge 0}  g_i(z) = m_i, i = \overline{1,m},  \sup\limits_{z_j \ge 0,j = \overline{1,m}}   g_0(z_1,\ldots,z_m) = m_0 \nonumber\\
	&& \sup\limits_{z \ge 0}J_0(z)= m_J.
	\label{3}
\end{eqnarray}
\end{itemize}

We assume that atmospheric pressure has the zero value. Since pressures can not be negative, the main goal of our paper is to find the conditions which guarantee the existence of bounded positive  solutions. Moreover, the coordinates of the solutions are separated from zero by some positive constants, and eventually the first coordinate must be larger than any another.  The solutions  should be  exponentially stable under certain conditions, and their various oscillatory properties (periodicity, eventually periodicity, almost periodicity)  correspondly to the particular properties of the sequence of moments $\theta_i.$ 
 
 Differential equations (\ref{1}) as a model of the physiological process are developed using models, which can be found, for example,  in \cite{bellman,cav,hop,lerma, ottesen} and many others.
To clarify that, we shall describe the two following models. 

In paper \cite{lerma} the following   differential equation is considered for the peripheric blood pressure $P_p$ 
\begin{eqnarray}
	&& \frac{dP_p}{dt} = -\frac{1}{RC}P_p + \frac{1}{C}Q(t),
	\label{c}
\end{eqnarray}
where   $C$ is  the total arterial compliance, $R-$ the arterial resistence, $Q(t)-$  the continuous blood flow. The formula for the continuous blood flow evaluation is given by 
$Q(t)= \frac{P_0(t)-P_p(t)}{r},$ where $P_0-$ is the systemic arterial pressure, and $r-$ is the aortic impedence.
As we suppose that the systemic arterial pressure is larger than the blood periferic pressure, $Q(t)$ is a positive function.

For the systemic arterial pressure   the following  differential equation with impulses has been discussed  in  book  \cite{hop},
\begin{eqnarray}
	&& \frac{dP_0}{dt} = - \frac{1}{RC} P_0, t \not = \theta_i,\nonumber\\ 
	&& P_0(\theta_i+) - P_0(\theta_i)= \frac{V_j}{C},
	\label{2}
\end{eqnarray}
where  $\theta_i= iT , i \in \mathbb Z,$ are prescribed moments,   $P_0(\theta_i+)$ is the right limit's value, $ V_j$ are the stroke volumes, and $C$ is the systemic arteria compliance.

 It is natural to  consider cardiovascular problems using discontinuous dynamics theory. One of the interesting approaches  which should be mentioned is the method of circle mappings \cite{arnold,glass} for cardiac arrhythmias. In \cite{akhmet1} we consider the extended form of  equation (\ref{2}) for the systemic arterial pressure,  and  we investigate the problem of existence of positive periodic and almost periodic solutions, oscillations of the equation. The case when the moments
of heart contraction are not prescribed, but are caused by a certain value of the systemic arterial pressure has also been investigated.

We have further developed the models of our predecessors in several ways:

 $1)$ The nonlinear perturbations caused by interactions are used in the model, both in differential equations and the impulsive part of the model. Function $g_0(P_0-P_1,P_0-P_2,\ldots,P_0-P_m)$ is positive since the outflow from the aorta to neighbouring  compartments reduces the systemic arterial pressure, all differences in the function are assumed to be positive. Nonlinearities $g_i(P_0-P_i), i=\overline{1,m},$ are positive, when the differences again are positive, since  the differences generate inflow into  compartments, which naturally  implies the rise of pressure. Every function  $g_i(P_0-P_i)$ does not depend on  $P_j, j \not =i,$ since we assume that  there is no essential mutual interaction between compatments;

  $2)$ The equations for the compartments are derived from equation (\ref{c}) for peripheric blood pressure. The nonnegative differences $\frac{P_0(t)-P_i(t)}{r_i}$ are replaced by the  nonlinear functions $g_i(P_0-P_i);$ 
  
  $3)$ $I_0$  is a constant part of the instanteneous change of systemic arterial pressure;

 $4)$ Condition $(C5)$ is chosen, assuming that the results obtained in our investigation
can be interpreted for the blood pressure system activated by an artificial pacemaker, or for the case when 
fluctuations are regular. That is, we exclude the case when the set of points of discontinuity may have so called accumulation points. 

We must also note that  our results can be easily reconsidered when condition $(C5)$ is replaced by condition 
$0< \mu \le \theta_{i+1} - \theta_i \le \mu_1 < \infty$ or by the condition  of existence of $\lim\sup_{T \to \infty} \frac{i(t_0,t_0+T)}{T} = p,$ where $i(t_0,t_0+T)$ is the number of elements  $\theta_i$ in the interval $ (t_0,t_0+T),$ and $p$ is a nonnegative real number, uniform for all $t_0 \in \mathbb R.$

The mathematical background of our  investigation, the theory of impulsive differential equations,  has been developed  in  \cite{akhmet7,aps,akhmet6,hw,laks,sp}.
The results of the paper concerning eventual periodicity, positiveness of solutions, domination of a coordinate over all other coordinates, are new. The role of the impulsive action as the one determining the relations is investigated for the  first time. We are aware that in reality the role of the impulses is diminished, for example, by delays,  and we plan to continue the investigation of  the  interactions of blood pressure levels in different parts of the cardiovascular system in our future papers. 

The paper is organized in the following manner. In the next section we find the conditions for the equation such that there exists a unique solution bounded on the whole $\mathbb R$ . The solution is periodic, almost periodic, or eventually periodic if the sequence of discontinuity moments has an appropriate property.    Section 3 contains the results on the stability and positiveness of the solutions. In the last section we specify the choice of the moments of discontinuity such that they may be determined artificially by a particular map, and by the initial moment. In this case the approach can be considered as a way to discuss the pacemaker's design.
Appropriate numerical simulations are provided.

 \section{Bounded, periodic, eventually periodic and almost periodic solutions}
 
 In this section we shall prove that the system admits oscillating solutions. The existence  of an eventually periodic solution is proved for the first time. The assertions about the  discontinuous periodic  and  almost periodic solutions are   specified versions of general theorems of the theory of impulsive differential equations \cite{aps,laks,sp}.

  Firstly, let us define the solutions of system (\ref{1}). 
    
  A left continuous function $P(t): R \to  R^{m+1}$ is in the set  ${\cal PC}( R),$ if  it is continuous on  $ R,$ except at the points $\theta_i, i \in  Z,$ where its first coordinate may have discontinuities  of the first kind, and  is left continuous.

A function $P(t) \in {\cal PC}( R)$ is a solution of (\ref{1}) if:
\begin{enumerate}
\item[(1)] the differential equation  is satisfied for $P(t), t \in  R,$ except at the points $\theta_i, 
i \in  Z,$  where it holds  for the left derivative of $P(t);$
\item[(2)] the jump equation  is satisfied by $P_0(t)$ for every   $ i \in  Z.$ 
\end{enumerate}
 
We shall also need the following set of functions.

 A left continuous function $P(t)$ is  in the set of functions ${\cal PC}( R_+), R_+ = [t_0, \infty), t_0 \in  R,$ if it  is continuous on $[t_0 , \infty)$ except at the points $\theta_i, t_0  \le \theta_i < \infty,$ where its first coordinate  may have discontinuities  of the first kind.

A solution $P(t) = (P_0(t),P_1(t),\ldots,P_m(t))$ of (\ref{1}) on $[t_0, \infty)$ is a function in
${\cal PC}( R_+)$ such that:
\begin{enumerate}
\item[(1)] the differential equation  is satisfied for $P(t)$ on $ R_+,$ except at the points $\theta_i, 
\theta_i > t_0,$  where it holds for the  left derivative of $P(t);$ 
\item[(3)] the jump equation  is satisfied by $P_0(t)$ for every   $ i, \theta_i > t_0.$ 
\end{enumerate}
 
 One can easily show, using the theory of impulsive differential equations \cite{laks,sp}, that under the aforementioned conditions a solution  $P(t,t_0,\pi_0), (t_0,\pi_0)  \in  R\times  R^{m+1}, \pi_0=(\pi_0^0,\pi_0^1,\ldots,\pi_0^m),$ of (\ref{1}) exists and is unique on $[t_0,\infty).$
 
 One can also find that the  solution satisfies the
 following integral equation 
 
 \begin{eqnarray}
&&P_0(t) =  {\rm e}^{-k_0(t-t_0)}\pi_0^0  -  \nonumber\\ && \int\limits_{t_0}^t{\rm e}^{-k_0(t-s)} g_0(P_0(s)-P_1(s),P_0(s)-P_2(s),\ldots,P_0(s)-P_m(s))ds + \nonumber\\
 &&\sum_{t_0 \le \theta_i < t}{\rm e}^{-k_0(t-\theta_i)}(I_0+J_0(P_0(\theta_i))),\nonumber\\
	&&P_i(t) =  {\rm e}^{-k_i(t-t_0)}\pi_0^i  +  \nonumber\\ && \int\limits_{t_0}^t{\rm e}^{-k_i(t-s)} g_i(P_0(s)-P_i(s))ds,  i= \overline{1,m}.
	\label{5}
\end{eqnarray}

We say that a sequence $\theta_i$ has the {\it $p-$property}, $p \in  N,$  if $\theta_{i+p} = \theta_{i} + p\omega$ for all integers $i.$ Moreover, we say that the sequence has the $p-$property eventually, 
if  $\theta_{i+p} - \theta_{i} \to p\omega$ as $i \to \infty.$

Consider a strictly ordered sequence of real numbers $t_i,i \in  Z.$  Denote $t_j^i = t_{i+j} - t_i, i,j \in  Z,$ and define  the sequences 
$\{t_j^i\}_i,i,j \in  Z.$ Following \cite{sp,hw} we call this family of sequences  equipotentially almost periodic if for an arbitrary positive $\epsilon$ there exists a relatively dense set of $\epsilon-$almost periods, common for all sequences $\{t_j^i\},j \in  Z.$

 It is proved in  \cite{aps}  (see also \cite{sp})  that the family $\theta_j^i, i,j \in  Z,$ is equipotentially almost periodic if the sequence $\theta_i - i \omega, i\in  Z,$ is almost periodic.
 That is, if we chose an almost periodic sequence $\xi_i,  i \in  Z,$ such that $0 \le  \xi_i < \omega,  i \in  Z,$ then the sequence $\theta_i,$ where $\theta_i = i \omega + \xi_i, i\in  Z,$ is equipotentially almost periodic.
 
 Let $\hat{[a,b]}$ be an oriented interval, that is  $\hat{[a,b]} = [a,b],$ if $a \le b,$ and 
  $\hat{[a,b]} = [b,a],$ if $a>b.$
   
 We say that a function $\phi(t)$ in ${\cal PC}( R)$ is eventually $p\omega-$periodic if the sequence 
 of discontinuities  $\theta_i$ has the $p-$property eventually, and 
 
\begin{eqnarray}\label{epf}
&& \lim_{t \to \infty}[\phi(t + p\omega)-\phi(t)] = 0,	
\end{eqnarray}
 for all $t$ such that $t \not \in \hat{[\theta_i,\theta_{i+p}-p\omega]}\cap\hat{[\theta_{i+1},\theta_{i+p+1}-p\omega]}.$

Next, we shall give a definition from \cite{sp,hw} which is slightly modified for our case.

A function $\phi(t)$ in ${\cal PC}( R)$ is a discontinuous almost periodic (d.a.p.) function if:
\begin{enumerate}
	\item [(a)] the function is uniformly continuous on the union of all intervals of the continuity $(\theta_i,\theta_{i+1}),  i\in  Z;$
		\item [(b)] for every positive $\epsilon$ there exists a relatively dense set $\Gamma$ of almost periods such that if $\gamma \in \Gamma,$ then $\|\phi(t +\gamma) -\phi(t)\| < \epsilon$ for all $t \in  R$ such that
		 $|t - \theta_i| > \epsilon, i\in  Z.$  
\end{enumerate} 

We may assume that
  	 \begin{itemize}
  	 	\item [(C6)]$ L(l_0,l_1,\ldots,l_m,l_J) = \sqrt{[\frac{l_0\sqrt{2m}}{k_0}+ \frac{l_J{\rm e}^{k_0\omega}}{1 -{\rm e}^{-k_0\omega}}]^2 + \sum\limits_{i=1}^{m}\frac{l_i^2}{k_i^2}}	< 1.$ 
	\end{itemize}

 \begin{theorem} \label{theorbd1}If conditions $(C1)-(C6)$ are fulfilled, 
then there exists a unique solution of (\ref{1}) bounded on $ R$ and:
\begin{enumerate}
	\item the bounded solution has the period $p \omega$ if the sequence $\theta_i$ has the $p-$property for a fixed $p \in  N;$ 
	\item  the bounded solution is an eventually $p\omega-$periodic function  if the sequence $\theta_i$ has the $p-$property eventually for a fixed $p \in  N;$ 
	\item  the bounded solution is a
d.a.p. function	if the sequence $\theta_i - i \omega, i\in  Z,$ is almost periodic. 
\end{enumerate} 
\end{theorem}

\textbf{Proof:} We shall use  the norm $|\phi|_0 = \sup_{ R}|\phi(t)|$ for  scalar valued  functions defined on 
$ R,$ and  $||\phi||_0 = \sup_{ R}\|\phi(t)\|$  for vector-valued functions.  Denote by $\Omega$ a subset of  ${\cal PC}( R)$  such that if $\phi(t) \in \Omega, \phi = (\phi_0,\phi_1,\ldots,\phi_m),$ then 
 $|\phi_0|_0 \le (\frac{m_0}{k_0}+ \frac{{m_J\rm e}^{k_0\omega}}{1 -{\rm e}^{-k_0\omega}}),\,
 |\phi_i|_0 \le \frac{m_i}{k_i}, i = \overline{1,m} .$
 
 One can easily verify that (\ref{1}) has a bounded solution $P(t)$  if and only if  $P(t)$  is  a bounded on $R$ solution of the following integral equation
  \begin{eqnarray}
&&P_0(t) = -\int\limits_{-\infty}^t{\rm e}^{-k_0(t-s)} g_0(P_0(s)-P_1(s),\ldots,P_0(s)-P_m(s))ds + \nonumber\\
 &&\sum_{ \theta_i < t}{\rm e}^{-k_0(t-\theta_i)}(I_0+J_0(P_0(\theta_i))),\nonumber\\
	&&P_i(t) =   \int\limits_{-\infty }^t{\rm e}^{-k_i(t-s)} g_i(P_0(s)-P_i(s))ds, i = \overline{1,m}.
	\label{6}
\end{eqnarray}

 Define on $\Omega$ an operator $T$ such that 
	\begin{eqnarray}
&&(T \phi)_0(t) =   -\int\limits_{-\infty}^t{\rm e}^{-k_0(t-s)} g_0(\phi_0(s)-\phi_1(s),\ldots,\phi_0(s)-\phi_m(s))ds + \nonumber\\
 &&\sum_{ \theta_i < t}{\rm e}^{-k_0(t-\theta_i)}(I_0+J_0(\phi_0(\theta_i))),\nonumber\\
	&&(T \phi)_i(t) =   \int\limits_{-\infty }^t{\rm e}^{-k_i(t-s)} g_i(\phi_0(s)-\phi_i(s))ds,  i= \overline{1,m}.
	\label{6,}
 \end{eqnarray}
We can verify  that $T:  \Omega \to \Omega.$ 
Take $\phi, \psi \in \Omega.$ 
One  can obtain  that  
	\[||T\phi -T\psi||_0 \leq  L(l_0,l_1,\ldots,l_m,l_J) ||\phi - \psi||_0,\]	 
and, consequently, condition $(C6)$ implies that  the operator is contractive.  It is not difficult to check that the space $\Omega$
	is complete. Thus, there exists a unique solution of (\ref{1}) from  $\Omega.$
	
	Assume now that the sequence $\theta_i$ has the  $p-$property.  Then, using the standard method, starting with a $p \omega-$periodic function from $\Omega$, we can, using the operator $T,$ construct a sequence of $p\omega-$periodic approximations of the bounded solution, which is also $p\omega-$periodic.
	
	 Consider the case where  $\theta_i$ has the  $p-$property eventually. In order to prove assertion $2)$ of the theorem,  it is sufficient, according to the above  discussion of boundedness of the solution,  to prove  that $T\phi$ is eventually $p\omega-$periodic, if  $\phi$ is an eventually $p\omega-$periodic function.  
	
	Fix a positive number $\epsilon.$ Since  $\theta_i$ has the  $p-$property eventually, and  $\phi$ is eventually  $p\omega-$periodic function,  one can find positive numbers $T_2>T_1$ such that: 
	
\begin{enumerate}
	\item  if $t \ge T_2$ and 
	$\theta_i+\epsilon < t < \theta_{i+1}-\epsilon$ for some $i \in  Z,$ then  $\theta_{i+p}<t + p\omega<\theta_{i+1+p}$ and 
	$\|\phi(t+p\omega) - \phi(t)\| < \epsilon; $ 
		\item  $2 {\rm e}^{-k_0(T_2-T_1)}[ \frac{m_0}{k_0}+   \frac{{m_J\rm e}^{k_0\omega}}{1 -{\rm e}^{-k_0\omega}}] < \epsilon, \,2{\rm e}^{-k_0(T_2-T_1)}\frac{2m_i}{k_i}< \epsilon.$	
\end{enumerate}

	We have that 
	\[|(T\phi)_0(t+ p\omega) -(T\phi)_0(t)| = |-\int_{-\infty}^{t+p\omega}{\rm e}^{-k_0(t+p\omega-s)} g_0(\phi_0(s)-\phi_1(s),\ldots,\phi_0(s)-\phi_m(s))ds  + \]\[ \sum_{\theta_i< t+p\omega }{\rm e}^{-k_0(t+p\omega-\theta_i)}(I_0+J_0(\phi_0(\theta_i)))+ \int_{-\infty}^t{\rm e}^{-k_0(t-s)}  g_0(\phi_0(s)-\phi_1(s),\ldots,\phi_0(s)-\phi_m(s))ds  - \]\[\sum_{\theta_i< t}{\rm e}^{-k_0(t-\theta_i)}(I_0+J_0(\phi_0(\theta_i)))| \le
		 \int_{-\infty}^{T_1}  e^{-k_0(t-s)}2m_0ds + \] \[ \sum_{\theta_i< T_1}e^{-k_0(t-\theta_i)}2m_J +
	 \int_{T_1}^{t}  e^{-k_0(t-s)}l_0\sqrt{2m}||\phi(s+p\omega) - \phi(s)||ds + \] \[ \sum_{T_1 \le \theta_i< t}e^{-k_0(t-\theta_i)}l_J||\phi(\theta_{i+p}) - \phi(\theta_i)|| + 
	 \sum_{\theta_i< t}\int_{\theta_i-\epsilon}^{\theta_i+\epsilon}  {\rm e} ^{-k_0(t-s)}2m_0ds \le\]\[	 	
\epsilon [1+ \frac{l_0\sqrt{2m}}{k_0}+ \frac{{\rm e}^{\kappa\omega}}{1 -{\rm e}^{-\kappa\omega}} (l_J + 4m_0{\rm e}^{k_0\epsilon})].\]	

Similarly, one can obtain that 
	\[|(T\phi)_i(t+ p\omega) -(T\phi)_i(t)| \le \epsilon [1+ \frac{l_i}{k_i} + 4m_i{\rm e}^{k_0\epsilon}\frac{{\rm e}^{\kappa\omega}}{1 -{\rm e}^{-\kappa\omega}} ],  i= \overline{1,m} .\]

That is, there exists a unique eventually $p\omega-$periodic solution of (\ref{1}).		
	
	 Now assume that the sequence $\theta_i - i\omega$ is almost periodic. From the previous discussion we see that to prove the  existence of a d.a.p. solution we need only to verify that the 
	function $T\phi$ is d.a.p.  if $\phi$ is a d.a.p. function.
	By Lemma 35 \cite{sp} (see also \cite{aps}),  for a given positive $\epsilon$ there exist 
	a real number $\nu, 0 < \nu < \epsilon,$ and relatively dense sets  of real numbers $\Gamma$ and integers $H$, such that: 
	
\begin{enumerate}
	\item $\|\phi(t+\gamma) - \phi(t)\| < \epsilon;$
	\item $|\theta_k^h - \gamma| < \nu, k \in  Z, h \in H, \gamma \in \Gamma.$
\end{enumerate}
Take the real number $\nu$ such that $\|\phi(t_1) - \phi(t_2)\| < \epsilon$ if $t_1,t_2$ belong to the same interval of continuity of the function $\phi(t)$ and $|t_1-t_2| <\nu.$
	We have that if $|t - \theta_i| > \epsilon, i\in  Z,$ then 
		\[|(T\phi)_0(t+\gamma) -(T\phi)_0(t)| = |\int_{-\infty}^{t+\gamma}{\rm e}^{-k_0(t+\gamma-s)}g_0(\phi_0(s)-\phi_1(s),\ldots,\phi_0(s)-\phi_m(s)) ds  + \]\[ \sum_{\theta_i< t+\gamma }{\rm e}^{-k_0(t+\gamma-\theta_i)}(I_0+J_0(\phi(\theta_i)))- \int_{-\infty}^t{\rm e}^{-k_0(t-s)}g_0(\phi_0(s)-\phi_1(s),\ldots,\phi_0(s)-\phi_m(s))ds  - \]\[\sum_{\theta_i< t}{\rm e}^{-k_0(t-\theta_i)}(I_0+J_0(\phi(\theta_i)))| \le 		  
	 \int_{-\infty}^{t}  e^{-k_0(t-s)}l_0\sqrt{2m}||\phi(s+\gamma) - \phi(s)||ds + \] \[ \sum_{\theta_i< t}e^{-k_0(t-\theta_i)}l_J||\phi(\theta_{i+h}) - \phi(\theta_i)|| + 
	 \sum_{\theta_i< t}\int_{\theta_i-\epsilon}^{\theta_i+\epsilon}  {\rm e} ^{-k_0(t-s)}2m_0ds \le\]\[	 	
\epsilon [\frac{l_0\sqrt{2m}}{k_0}+ \frac{2{\rm e}^{k_0\omega}}{1 -{\rm e}^{-k_0\omega}} (l_J + 4m_0{\rm e}^{k_0\epsilon})]\]
and 

\[|(T\phi)_i(t+ \gamma) -(T\phi)_i(t)| \le \epsilon [ \frac{l_i}{k_i} + 4m_i{\rm e}^{k_i\epsilon}\frac{{\rm e}^{k_i\omega}}{1 -{\rm e}^{-k_i\omega}}],  i= \overline{1,m} .\]

That is,  	$T\phi$ is a d.a.p. function.	
	The theorem is proved.
	
	\section{Stability and positiveness}
	
	In this section we  show that the bounded solution is exponentially stable, and that every coordinate of the solution is separated from zero by some positive number. Moreover, the first coordinate of the solution, which is the value of the systemic arterial pressure, is higher than any other pressure value. 
	
		Let us denote the solution bounded on $ R$ as 	 $\xi(t) = (\xi_0(t),\xi_1(t),\ldots,\xi_m(t)).$ Next, we find the conditions for the positiveness of this solution.
	Assume additionally that 
	
  	 \begin{itemize}
  	 	\item [(C8)]$ \frac{m_0}{k_0}+ \frac{m_i}{k_i} < I_0\frac{{\rm e}^{-k_0\omega}}{1 -{\rm e}^{-k_0\omega}},  i= \overline{1,m} .$ 
	\end{itemize}

	 Using (\ref{6}) we obtain that

	\[\xi_0(t) \ge I_0\frac{{\rm e}^{-k_0\omega}}{1 -{\rm e}^{-k_0\omega}} - \frac{m_0}{k_0} > 0, t \in  R,
\]
and 
	\[|\xi_i(t)| \le  \frac{m_i}{k_i}, t \in  R,i= \overline{1,m}.  
\]
Hence,
\begin{eqnarray} \label{comp}
	\xi_0(t) - \xi_i(t) \ge I_0\frac{{\rm e}^{-k_0\omega}}{1 -{\rm e}^{-k_0\omega}} - \frac{m_0}{k_0}-
\frac{m_i}{k_i} = \delta > 0, t \in  R, i= \overline{1,m},
\end{eqnarray}

and using   (\ref{6}) again one can see that 
	\[\xi_i(t) \ge  \frac{\bar g_i}{k_i} > 0, t \in  R, i= \overline{1,m},
\]
where $\bar g_i$ is the minimal value of the function $ g_i(z)$ for $ z \in [\delta,\frac{m_0}{k_0}+ \frac{{m_J\rm e}^{k_0\omega}}{1 -{\rm e}^{-k_0\omega}} + \frac{m_i}{k_i}].$

The following assertion is proved.

\begin{theorem}
Assume that  conditions $(C1)-(C8)$ are valid. Then for the  bounded  solution $\xi(t)$  of (\ref{1}) 
there exist positive constants $\nu_j,  j = \overline{1,m},\, \mu_i, i= \overline{0,m+1}$ such that the inequalities   $ \xi_0(t) - \xi_i(t) > \nu_j, \xi_i(t) \ge  \mu_i$ are valid for all $i,\,j,$ and  $t \in  R.$ 

\label{t7}
\end{theorem}

Let us  give the definition of  uniform exponential stability  
	of the solution. Denote $P(t) = P(t,t_0,P_0),$ a solution of (\ref{1}).

\begin{definition}   The solution   $\xi(t)$  is called uniformly exponentially stable if there exists a number 
$\alpha \in R, \alpha >0,$ such that for every $\epsilon > 0$ there exists a number $\delta = \delta (\epsilon)$ such that the inequality $||P(t)-\xi(t)|| < \epsilon \exp(-\alpha (t-t_0)), \forall t \geq t_0,$ holds, if
 $||P_0 - \xi(t_0)|| < \delta.$ 

\label{def4}
\end{definition} 
Fix a positive number  $\sigma, 0< \sigma <  \min_{i= \overline{0,m}}k_i,$ denote $m(l_0,l_1,\ldots,l_m,l_J) = 1 - \max\{\frac{ l_0\sqrt{2m}}{k_0 - \sigma} +  \frac{{l_J \rm e}^{(k_0-\sigma)\omega}}{1 -{\rm e}^{-(k_0-\sigma)\omega}}, \frac{ 2l_i}{k_i - \sigma}, i = \overline{0,m}\},$ and assume that the Lipschitz coefficients are  small so that
 \begin{itemize}
	\item [(C9)] $m(l_0,l_1,\ldots,l_m,l_J) > 0.$
	\end{itemize}

\begin{theorem}
Assume that  conditions $(C1)-(C7),(C9)$ are valid. Then the  bounded  solution $\xi(t)$  of (\ref{1}) 
is uniformly exponentially stable. 
\label{t3}
\end{theorem}

{\it Proof.} \rm   One can see that   $v(t) = P(t) - \xi(t), v = (v_0,v_1,\ldots,v_m),$ is  a solution of the equation 
\begin{eqnarray}
&& \frac{dv_0}{dt} = -k_0v_0 + w_0(v), \nonumber\\
&& \frac{dv_i}{dt} = -k_iv_i + w_i(v_0,v_i), i = \overline{1,m}, t \not = \theta_j,\nonumber\\
&&  \Delta v_0 |_{t=\theta_j} = u_0(v_0).	
\label{7}
\end{eqnarray}
where functions $w_i, i = \overline{0,m},$  and $u_0$ satisfy the following  conditions: 
\begin{eqnarray*}
&& |w_0(v)| \leq l_0\sqrt{2m} ||v||;\nonumber\\
&& |w_i(v_i,v_0)| \leq l_i (|v_i|+|v_0|), i \in  Z;\nonumber\\
&& |u_0(v_0)| \leq l_J|v_0|.
\end{eqnarray*}
Thus,  the problem of the stability of $\xi_0(t)$ is reduced to  the stability of  the zero solution
of (\ref{7}). Fix $\epsilon >0 $  and denote $ K=K(l_0,l_1,\ldots,l_m,l_J,\delta) = \frac{\delta}{m(l_0,l_1,\ldots,l_m,l_J)} , $ where $\delta \in R, \delta >0,$ and
take  $\delta$ so small
 that $ K(l_0,l_1,\ldots,l_m,l_J,\delta)<\epsilon.$
Assume, without loss of  generality, that $t_0 = 0.$  Let $v(t,v_0)$ be a solution of (\ref{7}) such that  $v(0,v_0)=v_0 =(v^0,v^1,\ldots,v^m).$

Denote by $\Psi$  a set of all functions  $\psi = (\psi_0,\psi_1,\psi_2,\ldots,\psi_m),$ defined on  $R_+ =[0, \infty),$ such that:  $1) \psi(0) = v_0;$ \quad $2)\,\psi(t) \in {\cal PC}( R_+);$ 
$3) \,  | \psi_i(t)| \leq K \exp(-\sigma t)$
 if $ t\geq 0, i = \overline{0,m}.$ Define  an operator $\Pi$ on $\Psi$ such that if $\psi \in \Psi,$ then

 \begin{eqnarray}
&&(\Pi \psi)_0=   {\rm e}^{-k_0 t}v^0  +   \int\limits_{0}^t{\rm e}^{-k_0(t-s)} w_0( \psi(s))ds + \sum_{0 \le \theta_i < t}{\rm e}^{-k_0(t-\theta_i)}u_0(v_0(\theta_i))),\nonumber\\
	&&(\Pi \psi)_i =   {\rm e}^{-k_i t}v^i  +   \int\limits_{0}^t{\rm e}^{-k_i(t-s)} w_i(v_0(s),v_i(s))ds,  i= \overline{1,m}.
	\label{8}
\end{eqnarray}
  We shall show that $\Pi : \Psi \rightarrow \Psi.$   
Indeed, for $ t\geq 0$ it is true that 
$$|(\Pi\psi)_0| \leq \exp(-k_0 t) \delta +  \int_{0}^t \exp(-k_0(t-s))l_0 \sqrt{2m}K  K \exp(-\sigma s)ds + $$$$\sum\limits_{0\le \theta_i<t}  {\rm e}^{-k_0(t- \theta_i)}l_J K
\exp(-\sigma \theta_i) \le \exp(-\sigma t)[\delta + K (\frac{ l_0\sqrt{2m}}{k_0 - \sigma} +  \frac{l_J{\rm e}^{(k_0-\sigma)\omega}}{1 -{\rm e}^{-(k_0-\sigma)\omega}})] \le K \exp(-\sigma t).$$

Similarly,

	\[|(\Pi\psi)_i| \leq \exp(-\sigma t)[\delta +2K \frac{ l_i}{k_i - \sigma}]\le K \exp(-\sigma t), i= \overline{1,m}.
\]
Let $\psi_1, \psi_2 \in \Psi_{\eta}.$
Then, we have that 
$$||\Pi\psi_1 -  \Pi\psi_2|| \leq  L(l_0,l_1,\ldots,l_m,l_J)\sup_{t\geq 0}||\psi_ 1- \psi_2||,$$
where the coefficient is described in $(C7).$
Using the contraction mapping argument  one can conclude that there exists 
a unique fixed point $v(t, v_0)$ of the operator $\Pi :\  \Psi \rightarrow \Psi$
which is a solution   of (\ref{7}). 
Theorem is proved.	

From (\ref{comp}) it follows that the bounded solution has the first coordinate larger than any other coordinate, and the attractiveness of the solution implies that any other solution in its neighbourhood  eventually has the first coordinate as its largest coordinate.   Now, we can  make some physiological conclusions: the normal state of the distribution of blood pressure is such that systemic arterial pressure is higher than  any other pressure. The initial state may be odd, that is,  the initial systemic arterial pressure may be lower than the pressure in a neighbouring compartment, but after a certain  period of time the state becomes normal. We may call this period  the time of stabilization of a solution. 

In order to carry out numerical simulations of the obtained theoretical results we  consider
the following equation

\begin{eqnarray}\label{exa1}
&&P_0' = -0.5P_0 - 0.1(P_0-P_1),\nonumber\\
	&&P_1' = -0.7P_1 + 0.1(P_0-P_1),\nonumber\\
	&&P_2' = -1.2P_1 + 0.1(P_0-P_2),\nonumber\\	
	&&\Delta P_0|_{t = \theta_i} = 0.07,
\end{eqnarray}
where $\theta_i = i + \frac{1}{4}|\sin(i) - \sin(\sqrt{2}i)|.$  It is proved in  \cite{aps} that the sequence $\frac{1}{4}|\sin(i) - \sin(\sqrt{2}i)|$ is almost periodic. The theoretical part of the paper implies
that system (\ref{exa1}) has a discontinuous almost periodic solution.

As  can be seen from Fig. 1, the solution  $P(t), P(0) = (0.08,0.05,0.03)$ is approaching the almost periodic solution  as time increases. 
\begin{figure}[hpbt]
 \begin{center}
  \epsfig {file=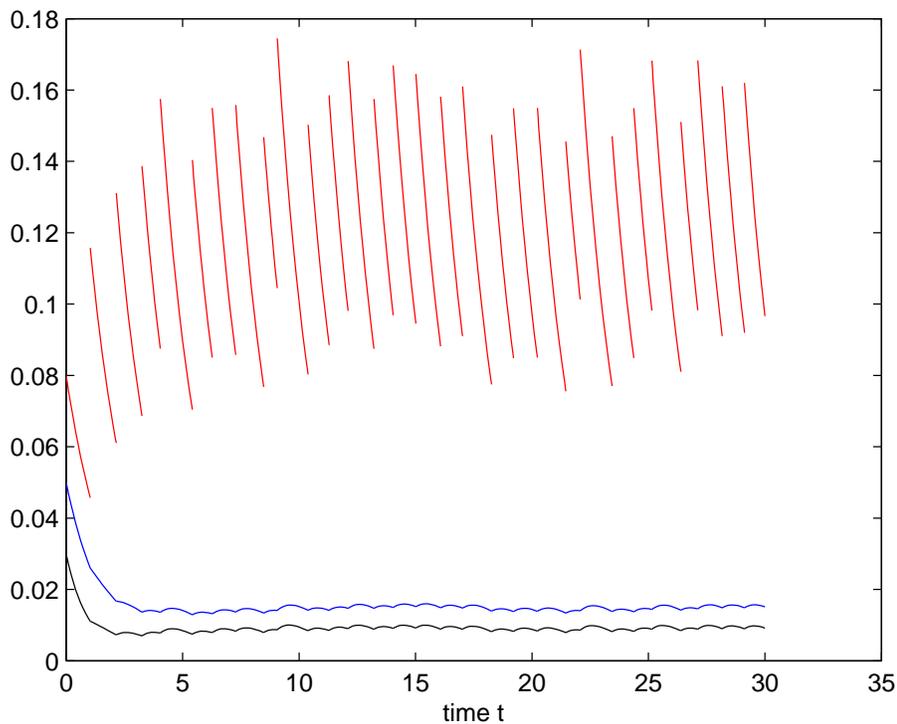,width=5.5in}
  \vspace{1.0cm}
  \caption{The graph of the first coordinate $P_0(t)$ of the solution is shown in  red, of  the second coordinate $P_1(t)$ in blue, and the third coordinate $P_2(t)$ in black. The initial value of the first coordinate  is larger than that of the second and third coordinates, and one can easily see from the graphs that our theoretical predictions are in full accordance with the properties represented by  the figure: the solutions are separated from  zero by some positive numbers,
coordinate $P_0(t)$ is always larger than  $P_1(t),P_2(t),$ and we may suppose that the solution is approaching the discontinuous  almost periodic solution of the system, as time increases.}
  \end{center}
\end{figure}

 \newpage

Fig. 2 shows that there exists a time of stabilization of a solution with  initial value $P(0) = (0.03,0.5,0.2)$ so that the first coordinate  is  smaller than  the second and third ones at the initial moment  $t=0.$ 
 
 \begin{figure}[hpbt]
 \begin{center}
  \epsfig {file=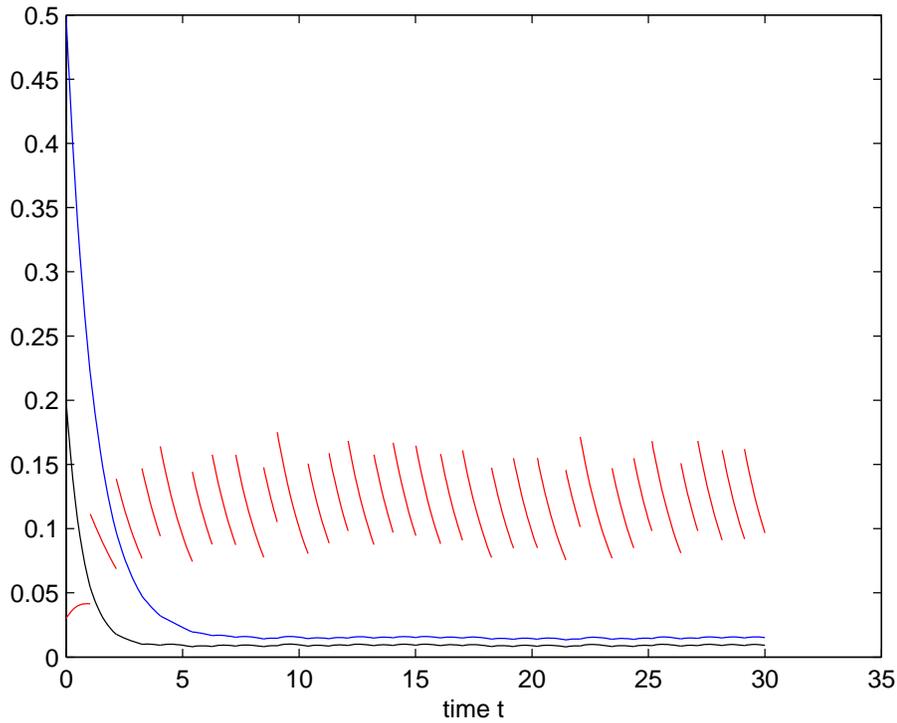,width=5.5in}
  \vspace{1.0cm}
  \caption{We can see  that despite the ``odd''  relation between the coordinates at the initial moment,    the normal state is eventually achieved, when coordinate $P_0(t)$ is larger than  $P_1(t)$ and  $P_2(t).$ The stabilization time of the solution  is no more than 5 units. Again, we have the consistency of the theoretical result with the simulation observations.}
  \end{center}
\end{figure}
\newpage

\section{Conclusion} As follows from the preceding discussion,    if  system 
(\ref{1})  satisfies certain conditions, then systemic arterial  pressure and  the blood pressure in compartments oscillate, remaining  positive. The shape of the oscillations depends on the behavior of the moments when   contraction of the left ventricle takes place. The oscillations are asymptotically stable. That is, they do not react significantly to external perturbations.  

From the results of our paper one can see that the strike type influence  of the  left ventricle contraction on the systemic arterial pressure  is ``softened'' when it reaches the peripheral compartments through the connections of the system, so that the compartments' oscillations are continuous. On the other hand, this influence is sufficiently large to  sustain the  positiveness of all coordinates and  keep  the aortic  pressure higher than in any other compartment.

  We believe that  the proposed  model can be developed further,  using experimental results as well as the  methods  of the theories of functional differential equations, discrete equations, etc,  to obtain additional features of the regular behavior of blood pressure, as well as to  investigate  irregular (chaotic) processes in the system. The problem of the period-doubling and intermittency routes to the chaos can serve as an important theoretical tool in the studies of the  arythmia processes.

\begin{Ack} The author thanks D. Alt\i ntan and C. B\"{u}y\"{u}kadal\i \,  for the technical assistance.
\end{Ack}

\end{document}